\documentclass{article} \font\cero=cmss10 scaled 1728
\font\uno=cmssbx10 scaled 1200 
\begin{document}
\begin{flushleft}
{\cero Fluctons} 
\end{flushleft} {\sf R. Cartas-Fuentevilla} and {\sf J.M. Solano-Altamirano}
\sf

{\it Instituto de F\'{\i}sica, Universidad Aut\'onoma de Puebla,

Apartado postal J-48 72570 Puebla Pue., M\'exico }\\

From the perspective of topological field theory we explore the
physics beyond instantons. We propose the {\it fluctons} as
nonperturbative topological fluctuations of vacuum, from which the
self-dual domain of instantons is attained as a particular case.
Invoking the Atiyah-Singer index theorem, we determine the
dimension of the corresponding {\it flucton moduli space}, which
gives the number of degrees of freedom of the {\it fluctons}. An
important consequence of these results is that the topological
phases of vacuum in non-Abelian gauge theories are not necessarily
associated with self-dual fields, but only with smooth fields.
Fluctons in different scenarios are considered, the basic aspects of the
quantum mechanical amplitude for fluctons are discussed, and the case of
gravity is discussed briefly. 

\vspace{1em}

\noindent {\uno I. Introduction}
\vspace{1em}

The discovery of instantons since the early seventies \cite{1} has
played historically an influential role in the understanding of
qualitative and quantitative aspects of non-Abelian gauge
theories. Specifically, they have revealed the very complicated
structure of the vacuum state of these theories, which is not
fully understood even at present. The nontrivial vacuum structure
has shown to have significant consequences on fundamental aspects
such as the strong CP violation \cite{2}, the $U(1)$ problem in QCD
\cite{3}, among others. Furthermore, in the mathematical context,
it is well known that instantons played a crucial role in the
discovery by Donaldson of simply connected 4-dimensional manifolds
with different smooth structures  \cite{4}. However, in the
confinement problem, the most intriguing aspect of QCD, these
nonperturbative vacuum fields turned out to be inapplicable in
constructing a confining mechanism. Additionally the realizations
of the self-duality condition associated with instantons require
very restrictive conditions on the base manifolds, {\it i.e.}
self-dual manifolds \cite{5}, manifolds with no two dimensional
anti-self-dual cohomology \cite{6}, noncompact gauge groups for
Lorentzian manifolds, etc. In fact, under certain conditions
there not exist instantons on
the $S^{2} \times S^{2}$ manifold \cite{4}.

In the search of physics beyond instantons, recently it has been
found that the so called Chern-Simons wave-function in the quantum
Yang-Mills theory represents actually ``topological excitations"
defined in the whole of the space of gauge connections of the
theory, and in particular in the self-dual subspace corresponding
to instantons\cite{7}. In \cite{7}, starting directly from the
topological Yang-Mills action, a symplectic structure is
constructed on the complete space of gauge connections, which
leads to a quantum Hamiltonian that admits the Chern-Simons
wave-function as an eigenstate with zero energy. In this manner,
neither the self-duality on the gauge connections nor the
Yang-Mills field equations are required in order to establish the
existence of the Chern-Simons state; the case corresponding to
instantons is attained as a particular case. In \cite{7} it is
elucidated on the possibility that an entirely similar situation
can be found in the context of loop quantum gravity for the Kodama
state starting directly from topological actions.

The physical reasons behind the results obtained in \cite{7} are
that if one starts from the ordinary Yang-Mills theory, the
self-duality condition on the gauge connections deforms the original
action into a topological one, obtaining thus the Chern-Simons
wave-function as a representation of a topological phase of the
theory of unbroken diffeomorphism invariance \cite{8}. In this
manner, the basic idea in \cite{7} was to start directly from a
topological action for Yang-Mills theory (which does not require
the self-dual connections for its definition), in order to
establish the results described above. This topological action has
as stationary points the complete space of smooth gauge
connections, and at the same time provides with a symplectic
structure to the corresponding phase space \cite{7}. Such a
geometric structure defined on the whole of the space of
connections allows to spread out the domain of definition of the
Chern-Simons state beyond the self-dual connections. Therefore,
one needs only a space of connections and a topological action
principle.

Let us call {\it fluctons} the topological excitations beyond the
self-dual domain of instantons. In this manner, the Chern-Simons
wave-function will represent the state with zero energy for
fluctons in concordance with the results in \cite{7}. In this
paper we shall determine the number of degrees of freedom of the
fluctons. In the next sections we outline the count of degrees of
freedom for instantons widely known in the literature, and we
shall extend the analysis to the case of fluctons.

This work is organized as follows. In the next section we give an
outline on ellipticity for instantons and its extension for
fluctons. In Section III, the relationship between instantons,
fluctons and unbroken topological phases is discussed. In Section
IV the elliptic complexes for instantons and fluctons are
considered. Invoking the Atiyah-Singer theorem, in Section V the
dimension of the {\it flucton moduli space} is determined.
Fluctons in different scenarios are considered in Section VI, 
a physical profile for fluctons is given in Section VII and
we finish in Section VIII with some concluding remarks. We follow
closely the notation and manipulations of
\cite{9}.

\noindent {\uno II. Ellipticity for instantons and fluctons}

\noindent {\uno IIa. Instantons}

Due to the gauge invariance of the theory, the Yang-Mills
equations are not elliptic as they stand, but a gauge fixing
condition is required. In order to impose the gauge condition, one
needs the theory of deformations of the space of solutions of the
Yang-Mills equations. Specifically the Yang-Mills equations
\begin{equation}
     d ^{\star} F + [A, ^{\star} F] = 0,
\end{equation}
together with the condition on the deformation $\delta A$ of the
gauge connection $A$
\begin{equation}
     d ^{\star} \delta A + [A, ^{\star} \delta A] = 0,
\end{equation}
constitute an elliptic system with only smooth solutions \cite{9}.
In this manner, when the gauge equivalence of solutions for the
Yang-Mills equations is taken into account, the corresponding {\it
moduli space} forms a finite dimensional space. This is true, in
particular, for instantons. Note that the ellipticity does not
depend on the self-dual character of instantons.

\noindent {\uno IIb. Fluctons}

We start, as mentioned in the introduction, directly from the
(finite) topological action for Yang-Mills theory,
\begin{equation}
     S_{TYM} (A) = \beta \int_{M} {\rm Tr} \ (F \wedge F),
\end{equation}
which does not depend on the metric on $M$ (a four-dimensional
manifold), however, it does on the smooth structure of $M$. $
\beta$ is a parameter. As well known, the action (3) implies the
Bianchi identities in a variational principle
\begin{equation}
     dF + [A,F] = 0,
\end{equation}
which is trivially satisfied since the Yang-Mills curvature comes
from a gauge connection
\begin{equation}
     F = dA + A \wedge A,
\end{equation}
{\it i.e.} every gauge connection $A$ is a critical point for this
action. If $D\equiv d + [A, \quad]$, we can understand Eqs.\ (4)
as the (non-linear) operator $DD$ with a kernel parametrised by
the complete space of gauge connections. Since the symmetry gauge
group $G$ acts (freely) on this space of connections $A$, we can
construct first the corresponding quotient space $A/G$; however,
if $A$ is a smooth connection satisfying the differential
identities (4), the gauge transformed connection
$g^{-1}Ag+g^{-1}dg$ is not necessarily smooth if $g$ is not
sufficiently smooth \cite{9}. Hence, we need to choose only the
smooth connections, and that is achieved fixing the gauge on this
space with the condition (2), and we shall obtain, by definition,
the {\it flucton moduli space}. Since the system of equations (2)
and (4) is elliptic (see Section (IVb)), this moduli space is
indeed finite dimensional. In this manner, we need to invoke
neither the Yang-Mills equations nor self-duality for obtaining an
elliptic system. The instanton scenario can be obtained as a
particular case, since the self-duality condition $F=\pm
^{\star}F$ will reduce the topological action (3) to the
conventional Yang-Mills action, the Bianchi identities (4) imply
the Yang-Mills equations (1), and the flucton moduli space reduces
to the instanton moduli space. In this sense, the flucton moduli
space must be considered as a finite dimensional subspace of the
infinity dimensional space $A/G$, and the instanton moduli space
as a subspace contained in the former. 

\noindent {\uno III. Unbroken topological invariance of the
action} 

\noindent {\uno IIIa. Instantons}

As mentioned in the introduction, in the case of Yang-Mills theory
the self-duality for instantons deforms the original action into a
topological one. The characterization of the tangent space of the
instanton moduli space is also through a self-dual condition on
the deformations of the curvature $\delta F = \pm ^{\star}\delta
F$, in order ''to maintain self-duality." However, in maintaining
self-duality actually one is keeping unbroken the topological
phase of the theory. Specifically at the level of the action, the
deformation of the curvature $F\rightarrow F+\delta F$ will lead
to the deformed action
\begin{eqnarray}
     \int_{M} {\rm Tr} (F \wedge ^{\star} F) \rightarrow \int_{M}
     {\rm Tr} \ (F + \delta F) \wedge^{\star} (F + \delta F),
\end{eqnarray}
which can be reduced to a topological action under the
self-duality conditions
\begin{eqnarray}
     F = \pm ^{\star}F, \quad \delta F = \pm ^{\star} \delta F,
\end{eqnarray}
hence, physics of instantons means, in this sense, physics within
topological phases.

\noindent {\uno IIIb. Fluctons}

If one starts from the topological action $\int_{M} {\rm Tr}  (F
\wedge F)$, and considers the deformation theory of the flucton
moduli space defined above, we need neither self-dual connections
on the background nor self-dual deformations in order to maintain
unbroken the topological invariance of the action, since the
deformed action
\begin{eqnarray}
     \int_{M} {\rm Tr}  (F \wedge F)\rightarrow \int_{M} {\rm Tr} \
     (F + \delta F)\wedge (F + \delta F),
\end{eqnarray}
is topological in character as the original one. In this manner,
the only condition  on the tangent space of the flucton moduli
space is given by Eq.\ (2), which in fact invokes the metric
structure on $M$, as an ordinary feature of topological field
theory. However, the fact of starting from a manifestly
topological action allows us to consider the following particular
cases: fluctons in self-dual backgrounds and arbitrary
deformations; fluctons in arbitrary backgrounds and self-dual
deformations; and finally the case of both self-dual backgrounds
and self-dual deformations, which will reduce fluctons to
instantons. In the present work we are considering the more
general situation, fluctons in arbitrary background and
deformation fields. Note that in all these cases
the topological invariance of the action is maintained unbroken. 

\noindent {\uno IV. Elliptic complex for instantons and fluctons}

\noindent {\uno IVa. Instantons}

The elliptic complex characterizing the deformation theory of the
instanton moduli space is given by \cite{9}
\begin{equation}
     0 \longrightarrow \Omega^{0}(M) \stackrel{D}{\longrightarrow} \Omega^{1}(M)
     \stackrel{D^-}{\longrightarrow} \Omega^{2}_{-}(M) \stackrel{\pi_+}{\longrightarrow} 0,
\end{equation}
where $\Omega^{i}(M)$ are the $i$-forms on $M$, $\pi_+$ ($\pi_{-}$) being the operator which projects a
2-form onto its self-dual (anti-self-dual) part and $D^{-} \equiv
\pi_{-}\circ D$. Of course, in the complex (9)
we have $ D^{-}\circ D=0.$

 The index of the complex is given by the alternating sum
\begin{equation}
     h_{0} - h_{1} + h_{2},
\end{equation}
where $h_{0} = dim (Ker D)$, $h_{1} = dim(Ker D^{-}/Im D)$, and
$h_{2} = dim(Ker \pi_{+}/Im D^{-})$. $h_{1}$ corresponds to the
dimension of the instanton moduli space ${\it M}_{k}$, $ h_{1} =
dim {\it M}_{k}$. Invoking the Atiyah-Singer theorem, and
considering that $h_{0} = 0 = h_{2}$ \cite{9}, one finds that
\begin{equation}\label{eq:dimmk}
     dim {\it M}_{k} = p_{1}(ad_{c} P) - \frac{dim G}{2} ( \chi(M) - \tau(M)),
\end{equation}
where $\chi(M)$, and $\tau (M)$ correspond to the Euler
characteristic, and the topological signature of $\it M$ respectively. $p_{1}$
corresponds to the first Pontrjagin class of the complexification
$ad_{c} P$ of the adjoint bundle $ad P$ of the instanton. $k$
corresponds to the second Chern class (see Eq.\ (3)), and
characterizes the equivalence classes of the instanton bundles.

It is important to mention that the condition $h_{2}=0$, essential
for determining $h_{1}$ by means of Eq.\ (11), requires  a
vanishing theorem, which in turn requires that $M$ is self-dual
with positive scalar curvature; these restrictions on $M$ will be
not invoked in the case of fluctons.

 In the case of $SU(2)$ instantons on $S^{4}$,
Eq.\ (11) reduces to the very well known expression
\begin{equation}
     dim {\it M}_{k} = 8 \kappa - 3,
\end{equation}
which is also valid for $SU(2)$ instantons on a simply connected
manifold with positive definite intersection form \cite{9}.

\noindent {\uno IVb. Fluctons}

Considering that the tangent space to the flucton moduli space is
characterized by Eq.\ (2), and that the deformations which differ
by a gauge transformation are equivalent ({\it i.e.} $\delta A$
and $\delta A + D\chi$ with $\chi\in \Omega^{0}$, are equivalent),
the natural elliptic complex for fluctons will be given by
\begin{equation}
     0 \longrightarrow \Omega^{0}(M) \stackrel{D}{\longrightarrow} \Omega^{1}(M)
     \stackrel{D^*}{\longrightarrow} 0,
\end{equation}
where it is evident that, by construction,
\begin{equation}
     D^{*}\circ D=0.
\end{equation}
The cohomological data for the flucton complex is $h_{0} = dim(Ker
D)$, and $h = dim(Ker D^{*}/Im D)$, and the corresponding index
will be the alternating sum
\begin{equation}
     h_{0} - h.
\end{equation}
Note that $h_{0}$ is exactly the same for the flucton and for the
instanton complex; hence $h_{0}=0$ in (15), since it is the
dimension of a space of sections \cite{9}. $h$ corresponds, of
course, to the dimension of the flucton moduli space $N_{k}$, $h =
dim N_{k}$. $k$ corresponds again to the second Chern class (see
Eq.\ (3)), and characterizes now the equivalence classes of the
flucton bundles.

Making a comparison, we may consider that the flucton complex (13)
is obtained by ''rolling up" the instanton complex (9), since does
not require self-duality; and conversely, that the later is
obtained by ''unrolling" the former since does require
self-duality.

Furthermore, considering that the scalar product on the involved fibres
is the usual one, it is easy to prove that the corresponding adjoint
operators are
\begin{equation}
     D^{\dag} = - D^{*}, \quad (D^{*})^{\dag} = - D,
\end{equation}
where we are considering that $D^{*}$ in the complex (13) acts on
1-forms. Hence, the relevant {\it Laplacians} of the flucton
complex are $\Delta_{0} \equiv D^{\dag} D + DD^{\dag}$, and
$\Delta_{1} = (D^*)^{\dag}D^{*} + D^{*}(D^*)^{\dag}$, which reduce to
\begin{equation}
     \Delta_{0} = D^{\dag}D + DD^{\dag} = - D^{*}D = \Delta_{1} =
     \Delta,
\end{equation}
taking into account Eqs.\ (16). Since we are considering a
Riemannian manifold, $\Delta$ is evidently an elliptic operator,
and therefore has a finite-dimensional kernel. In this manner, the
tangent space and the flucton moduli space itself are indeed
finite dimensional, and $h$ in Eq.\ (15) is well defined.

\noindent {\uno V Index theory and the dimension of the flucton
moduli space} 

In this manner, invoking the Atiyah-Singer theorem, and
considering that $h_{0} = 0$, we have an expression for the
dimension of the flucton moduli space
\begin{equation}
     h = dim N_{k} = ch (ad_{c} P) \overline{E} (M) [ M],
\end{equation}
where
\begin{equation}
     \overline{E} (M) = \frac{ch (\Sigma_{p} (-1)^{p}
     [\overline{E}^{p}])}{e(M)} td ( {\it TM}_{c}),
\end{equation}
and $\overline{E}^{0} = ad_cP\otimes\wedge^{0} T^{*}M_{c}$, $\overline{E}^{1}
= ad_cP\otimes\wedge^{1} T^{*}M_{c}$; $ad_{c}P$ stands for the
complexification of the adjoint bundle $ad P$ of the flucton, {\it
ch} for the corresponding Chern character, ${\it td}$ for the Todd
class, and ${\it e}$ for the Euler class.

Following \cite{9}, we can find that
\begin{equation}
     ch(ad_{c}P) = dim G + \frac{1}{2} p_{1} (ad_{c} P),
\end{equation}
as a consequence that ${\it M}$ is a four-dimensional manifold,
and that $ad_{c} P$ is the complexification of the real bundle;
thus, $ch(ad_{c}P)$ is the same for fluctons and for instantons.
The difference will be found in $\overline{E}(M)$. Following again
\cite{9}, we can see that the flucton complex (13) is a truncation
of the de Rham complex, with an index given by
\begin{equation}
     b_{0} - b_{1},
\end{equation}
where $b_{i}$ are the Betti numbers of ${\it M}$, in terms of
which the Euler characteristic of ${\it M}$ is given by
\begin{equation}
     \chi (M) = 2 b_{0} - 2 b_{1} + b_{2},
\end{equation}
considering the Poincar\'e duality; thus the four-dimensional
contribution for $\overline{E}(M)$ is
\begin{equation}
     b_{0} - b_{1} = \frac{1}{2} (\chi (M) - b_{2});
\end{equation}
therefore, taking into account Eqs.\ (18), (20), and (23), we find
for the dimension of $N_{k}$ the four-dimensional contribution of
the expression
\begin{equation}
     [dim G + \frac{1}{2} p_{1} (ad_{c} P)] [2 - \frac{1}{2}
     (\chi (M) - b_{2})],
\end{equation}
we mean
\begin{equation}\label{eq:dimflucmodsp}
     dim N_{k} = p_{1} (ad_{c} P) - \frac{1}{2} dim G (\chi (M)
     - b_{2}),
\end{equation}
which can be expressed in terms of the topological signature $\tau
(M) = b^{+}_{2} - b^{-}_{2}$, in order to make a direct comparison
with the corresponding expression for instantons:
\begin{equation}
     dim N_{k} = p_{1} (ad_{c} P) - \frac{1}{2} dim G [\chi (M)
     - \tau (M)] + b^{-}_{2} dim G.
\end{equation}
If the manifold $M$ is self-dual with positive scalar curvature,
then the first two terms correspond to the dimension of the
instanton moduli space (see Eq.\ (11)), and then
\begin{equation}
     dim N_{k} = dim M_{k} + b^{-}_{2} dim G;
\end{equation}
since $b^{-}_{2} dim G \geq 0$, it is evident in this case that
$M_{k}$ is indeed a subspace of $N_{k}$.

Equation (26) gives thus an expression for the flucton moduli
space in terms of topological invariants of the manifold ${\it M}$
and the fibre bundles involved. However, unlike the instanton
case, there are no res\-trictions on the self-dual character and
the scalar curvature of $M$. Let us consider now
some particular scenarios. 

\noindent {\uno VI. Examples}


 \noindent {\uno a) SU(2) fluctons on $S^{4}$}

In this case $b^{-}_{2} = 0$, and hence
\begin{equation}
     dim N_{k} = dim M_{k} = 8k -3,
\end{equation}
since for a $SU(2)$ bundle $\it E$ which gives
\begin{equation}
     k = - c_{2} (E),
\end{equation}
we have \cite{9}
\begin{equation}\label{eq:pontrclass}
     p_{1} (ad_{c} P) = -8 c_{2} (E) = 8k,
\end{equation}
in dimension four; thus, we can call $k$ in (28) as the {\it
flucton number}, which corresponds to the topological action (3)
(after a convenient normalization), and characterizes the topology
of the $SU(2)\!$ flucton bundles.

Note that the equality (28) does not imply that the fluctons
reduce to instantons, rather we do not require the Yang-Mills
equations and self-dual fields for establishing the existence of
nonperturbative topological vacuum fields for $SU(2)$ bundles on
$S^{4}$. If the starting point was those equations on $S^{4}$,
then their topological phases can be attained using self-duality,
and the resulting instantons turn out to have the same number of
degrees of freedom of those pseudo-particles living {\it de perse}
in the topological phases, the fluctons.

In the case of $SU(2)$ fluctons on a simply connected manifold with
positive definite intersection form such as the projective space
$CP^{2}$, we have also $b^{-}_{2} = 0$, and the same results and
interpretation given above are valid.

It is remarkable that although the elliptic complex (9) and (13)
are intrinsically different to each other, give exactly the same
dimension for the moduli spaces. However, this situation will
change in the case of $S^2\times S^2$ and $K3$ manifolds considered below.

\noindent {\uno b) SU(2) fluctons on $K3$ manifolds}

As well known $K3$ manifolds have played historically a crucial
role in string compactification, and will give also in the present
treatment a scenario where we shall find fundamental differences
between instantons and fluctons.
%
%

For this case 
we have 
$\chi=24$ and $b_2=22$ as the relevant topological invariants 
of $K3$; and by similar arguments to the previous examples,
$p_{1} (ad_{c} P) = -8 c_{2} (E) = 8k$ for the $SU(2)$ bundles. 
Thus, making use of the expression (\ref{eq:dimflucmodsp}), it is easy to see that
\begin{equation}\label{eq:dimmodspk3}
  dim N_{k} = 8k -3,
\end{equation}
which is a remarkable result due to the coincidence with the other results
obtained previously.

On the other hand, if we try to determine the dimension of the instanton
moduli space, we face the problem that there is not vanishing theorem
because the $K3$ manifolds have a 
non-zero anti-self-dual part of the Weyl tensor \cite{11}, which is a crucial requirement
for the vanishing theorem to be valid. Thus we can not use the expression (\ref{eq:dimmk})
in order to obtain the dimension of the instanton moduli space.

Again, one can observe the existence of nonperturbative topological vacuum fields
for $SU(2)$ bundles over $K3$ manifolds and furthermore, one can determine the 
dimension or their moduli space. This is not true for the instantons that,
although they can exist over $K3$ manifolds \cite{4}, one can not determine the dimension of
their moduli space. Due to this fact, the importance of the result (\ref{eq:dimmodspk3}) is 
considerably enhanced.

\noindent {\uno c) SU(2) fluctons on $S^2\times S^2$}

Since $\chi(S^2\times S^2)=\left(\chi(S^2)\right)^2$ we have $\chi=4$. On the other hand, 
we can determine $b_2$ by means of 
K\"unneth's formula obtaining $b_2=2$. Finally, following closely the 
arguments given in \cite{9}, we can find that $p_{1} (ad_{c} P) =8k$. Thus the dimension of the
flucton moduli space is determined by Eq. (\ref{eq:dimflucmodsp}) yielding 
\begin{equation}
   dim N_{k} = 8k -3,
\end{equation}
a result that we have found throughout this work.  
The most outstanding point of this case is that if $k=1$ then the instanton moduli
space is empty\footnote{In \cite{4}, this is equivalent to
$c_2=1$, where Donaldson analyses anti-self-dual connections instead of
self-dual connections;  and because of this, he deduces that $p_1=+8c_2=8k$  instead of $p_1=-8c_2=8k$ as we 
did in this treatment.}
\cite{4} while the flucton moduli space is not,
which means that, in this particular case,  nonperturbative topological vacuum fields
for $SU(2)$ bundles exist over the $S^2\times S^2$ manifold albeit instantons do not. 


\noindent {\uno VII. Physical profile of a flucton}

In this Section, we shall discuss briefly a possible physical picture of the fluctons.
As a first topic, the  quantum mechanical amplitude must be given by
\begin{equation}
   \exp[-S_T(A)]=\exp\left[-\frac{|k|}{g^2}\right]=
   \exp\left[-\frac{1}{8\pi^2g^2}\left|\int_MTr(F\wedge F)\right|\right]
\end{equation}
where $g$ is the coupling constant which had been considered equal to unity
until now, and $k$ is the flucton number.  In analogy with the instanton case \cite{12},
this amplitude is an inverse power series in $g^2$, and it is obvious that $k$
must be non-zero for the power series of the amplitude to exist. This amplitude 
could be interpreted as the transition amplitude between homotopically different
vacua. The advantage of our approach respect to the instanton analogue,
is that fluctons could describe this
transition between vacua even when we can not define (anti-)self-dual field
strengths. For instance, when the background manifold is $S^2\times S^2$, because it has
an empty moduli space, as we have seen before. 
One remaining problem is to determine the explicit form of the fluctons; but in principle, any
smooth connection is a flucton.

Let us now analize how the fluctons are related with gravity.  For arising
from a topological action, the fluctons have a zero energy-momentum tensor, and 
according to the Einstein's equation they do not produce curvature of
space-time and will not have gravitational effects
(of course, 
this result does not prohibit fluctons to interact with any other field
defined over the base manifold.) This is a property that fluctons share with instantons, since 
following the same procedure, one can conclude that the instantons
do not have any gravitational effect. The last is a consequence of the fact that instantons ``live''
in the topological sector of the theory (see Section IIIa).


 \noindent {\uno VIII. Concluding remarks} 

 In instanton theory the self-duality fixes the
infinite degrees of freedom of the space of gauge connections to
finite degrees of freedom given by the dimension of the moduli
space. In the case of fluctons treated here, the condition (2)
that demands only smooth fields is sufficient for fixing the
original infinite degrees of freedom of the space of gauge
connections to finite degrees of freedom for smooth fields, which
constitute thus the more general nonperturbative topological
fluctuations of the vacuum. In general, the fluctons will have a
contribution coming from the space of instantons.

On the other hand, the structure of the partition function of
topological field theory depends sensitively on the dimension of
the moduli space \cite{10}. In this manner, the additional term in
the expression (27) for the dimension of the flucton moduli space
in relation to that of the instanton can have important
consequences on the path integral formulation of the theory, and
subsequently on the identification of the corresponding partition
function as a topological invariant; it is by this way that the
celebrated Donaldson invariants are constructed from topological
quantum field theory invoking instantons.

Since self-duality restricts severely the properties of the base
manifolds, we have shown that the fluctons can exist in more general
manifolds where the instantons do not. Finally,
it is possible that following the arguments given in the present
treatment for TYM theory, similar results can be obtained in the
context of topological gravity, however this will be a
problem for future works. 

\begin{center}
{\uno ACKNOWLEDGMENTS}
\end{center}

This work was supported by the Sistema
Nacional de Investigadores and Conacyt (M\'{e}xico). 
The authors sincerely thank the useful comments of
G. F. Torres del Castillo.


\begin{thebibliography}{}
\setlength{\itemsep}{-.50em}
\bibitem{1} A. Belavin, A. Polyakov, A. Schwarz, and Yu. S.
Tyupkin, Phys. Lett. B. 59, 85 (1975).
\bibitem{2} C. Callan, R. Dashen, and D. Gross, Phys. Lett. B. 63, 334
(1976); R. Jackiw, and C. Rebbi, Phys. Rev. Lett. 37, 172 (1976).
\bibitem{3} G. 't Hooft, Phys. Rev. Lett. 37, 8 (1976); Phys. Rev.
D. 14, 3432 (1976).
\bibitem{4} S. Donaldson, P. Kronheimer, {\it The geometry of
four-manifolds}, Oxford University Press, Oxford (1990); D. Freed,
K. Uhlenbeck, {\it Instantons and four-manifolds}, 2nd ed.,
Springer-Verlag, New York (1988).
\bibitem{5} M. F. Atiyah, N. J. Hitchin, I. M. Singer, Proc. Roy.
Soc. Lond. A362, 425 (1978).
\bibitem{6} C. H. Taubes, J. Diff. Geom. 17, 139 (1982).
\bibitem{7} R. Cartas-Fuentevilla, and F. Tlapanco-Lim\'on, Phys.
Lett. B., (2005).
\bibitem{8} I. Oda, hep-th/0311149.
\bibitem{9} C. Nash, {\it Differential topology and quantum field
theory}, London Academic (1991); M. F. Atiyah,and I. M. Singer,
Bull. Amer. Math. Soc. 69, 422 (1963); Ann. Math., 87, 546 (1968).
\bibitem{11} T. Eguchi, P. B. Gilkey and A. J. Hanson, Phys. Rep,
{\bf 66}, 213 (1980).
\bibitem{10} E.~Witten, Comm.\ Math.\ Phys . {\bf 117}, 353
(1988).
\bibitem{12} C.~Nash,
  arXiv:hep-th/9709135, $\S 4$
\end{thebibliography}
\end{document}